\documentclass [10pt]{article}
\usepackage{graphics}
\begin{document}
\small{Subject classification: 71.23An; 71.30$+$h; 71.23$-$k

\vspace{2 mm}
\Large{\textbf{The delocalization problem of two interacting electrons
in a two-dimensional random potential}}
\normalsize

\vspace{2 mm}
\textsc{J. Talamantes}\footnote{Fax: +1 661 664 2040. email:
jtalamantes@csub.edu} \textit{(a)},
\textsc{M. Pollak} \textit{(b)} and \textsc{I. Varga}\footnote{permanent
address: Institute of Physics, Technical University of Budapest, Hungary}
\textit{(c)}\\
\small
\textit{(a) Department of Physics, California State University,
Bakersfield, CA 93311 USA\\
(b) Department of Physics, University of California,
Riverside, CA 92651 USA\\
(c) Fachbereich Physik, Philipps-Universit\"{a}t Marburg, Germany.}

\vspace{2 mm}
%
%
%
%
%
%
\small
The problem of two electrons in a two-dimensional random potential is
addressed numerically. Specifically, the role of the Coulomb interaction
between electrons on localization is investigated by writing the
Hamiltonian on a localized basis and diagonalizing it exactly. The result of
that procedure is discussed in terms of level statistics, and the expectation
value of the electron-electron separation.

\vspace{2 mm}
\normalsize
\textbf{Introduction.}
The effect of the Coulomb electron-electron 
interaction (EEI) on electronic localization
in a random potential is investigated by computer simulation
for a system with two electrons, using level
spacing statistics. In the absence of interactions, it has been shown (see \textit{e.g.}
\cite{Shklovskii&al}) that the
distribution shifts from Poisson 
to Wigner as the system goes from being strongly localized to delocalized. For \textit{interacting}
systems, such correspondence has never been proven. Still, level spacing
statistics has been used often to assess
localization. Here, we attempt to make 
a connection between localization and level spacing statistics for interacting
systems.

The combined importance of disorder, interaction, and elastic tunneling 
poses a very difficult problem. It was suggested twenty years ago
that the EEI can delocalize the electrons \cite{Pollak}, but as yet a firm
answer is lacking. Computationally, the main difficulty is the huge
phase space for systems of reasonable
size \cite{TPE}. Existing work (\cite{TPE} - \cite{Shep2}) resorted to various approximations.
The 2-electron problem for reasonably large systems
can be solved without such
approximations, double occupation of sites can be 
accounted for, spin
and exchange included, and the entire phase space can be examined.
The motivation for the 
problem considered here
is the ability to study the legitimacy of the approximations made in
the finite-density works, and thus shed some light on 
this more complex problem. We hope some insight
can be gained into the mechanisms at play in the experimentally observed
metal-insulator transition in 2{\it D\/} \cite{2dMIT:experiments}.
Previous relevant studies on the 2\textit{D} random system with interactions
include \cite{TPE}, finite-size scaling of three and four \cite{BS},
and two \cite{Emilio,Shep2} spinless electrons.
All these works have concluded that the
interaction enhances delocalization. In \cite{BS} a \textit{crossover}
from Poisson to Wigner was found, while \cite{Emilio,Shep2} reported a 
sharp transition. 

\vspace{2 mm}
\textbf{Computations.}
The procedure used here differs from methods used in
\cite{BS,Emilio,Shep2}
in that it is virtually exact in a tight binding scheme: spin is not ignored, fluctuations in the overlap
integrals are not neglected -- nearest-neighbor (\textit{n-n}), next \textit{n-n},
and next-to-next \textit{n-n} coherent tunneling processes are accounted for.
The method we use is the following. First, singlet and
triplet configurations are written for every pair of
sites in the system.  These configurations are constructed from one-electron
\textit{s}-orbitals $\varphi$ of radius $a_B$ centered
on the sites. The Hamiltonian is written
in this representation and diagonalized. The resulting eigenstates
are analyzed, and the level spacing distribution is computed for several
system sizes. 
The basis set used are the two-electron configurations $\phi_{ab}$ for a
given pair of sites $a$ and $b$. The $\phi_{ab}$ are constructed
by symmetrizing or antisymmetrizing products
such as $\varphi_{a}(1) \varphi_{b}(2)$ of one-electron orbitals for
electrons 1 and 2. We consider the Hamiltonian
\begin{equation}
H=\sum_{\alpha=1}^{2} (T_\alpha + V_\alpha+\varepsilon_\alpha)+
  \frac {e^{2}}{\kappa r_{12}}, \label{hamiltonian}
\end{equation}
where $\alpha$ labels the electrons, $T$, $V$, and $\varepsilon$ are the
operators for the kinetic energy,
the core potentials, and the random potentials, respectively;
$e$ is the electronic charge, $\kappa$ the dielectric constant,
and $r_{12}$ the distance between the
electrons. $\varepsilon_{\alpha}$ is chosen
from a box distribution $-W/2 \leq \varepsilon \leq W/2$, with $W$ equal
to the \textit{n-n} Coulomb energy.

The integrals 
corresponding to the matrix elements are performed numerically. An $L \times L$
lattice is set up, and $H$ is diagonalized for the parameters $L$ and
$r_s$ (the \textit{n-n} distance in units of $a_B$).
The resulting eigenstates $I$, and eigenenergies $E_I$ are investigated
in two ways: \textit{(i)} the distribution $p(s)$ of nearest-neighbor
level spacings is obtained. (We dropped 100 states or so from the band edges.)
As in \cite{Emilio}, a parameter
$\eta=(\mathrm{var}(s)-0.273)/(1.0-0.273)$ is computed as a measure of how close
$p(s)$ is to a Poisson ($\eta=1$) or Wigner ($\eta=0$) distribution; 
\textit{(ii)} the expectation value
$\lambda_{I} = \sum_{ab} r_{ab} |A_{I,ab}|^2$
of the \textit{e-e} separation
is computed from the eigenstates of (\ref{hamiltonian}). 
$r_{ab}$ is the distance (in units of $a_B$) 
between $a$ and $b$, and $A_{I,ab}$ are the coefficients defined by
$I=\sum_{ab} A_{I,ab} \phi_{ab}$.
One expects that
in the localized regime $\lambda_I$ is strongly correlated with $E_I$ (larger
$\lambda_I$ correspond to smaller $E_I$), whereas $\lambda_I$ should become
essentially independent of $E_I$ as configuration mixing increases.

For definiteness, $a_B$ and $\kappa$ are taken here to be 10\AA\ and 3
respectively -- these values seem appropriate for 2\textit{D} systems.
This choice of parameters yields an effective mass $m^*=0.16m$, with
$m$ the electron mass. The charge on the sites is taken to be $|e|$.
Cyclic boundary conditions are used.

\vspace{2 mm}
\textbf{Results and discussion.}
Runs were performed for $L=4,6,7,10$ and $r_{s}=5,6,7,8,9,10,11,12$. In every
case the number of ``samples'' was sufficient to obtain
$\sim 1.5 \times 10^4$ levels. Fig. 1 presents $\eta (r_s)$
for each value of $L$; furthermore, to show that the clear drop in $\eta$
with decreasing $r_s$ is due to the EEI (and not simply weak
localization due to system size), fig. 1 also shows $\eta$
without the last term in eq. (\ref{hamiltonian}). It is noted
that there is no clear small-size scaling behavior, suggesting
a crossover ($r_s \sim 9 \textrm{-} 11$),
rather than a transition. This is in agreement with 
\cite{BS}, but differs from \cite{Emilio,Shep2}.
The differences may be due to differences in models
and choice of parameters.

Fig. 2 presents fits for $\lambda (E)$ with $L=7$.
It represents a single realization of the disorder, but
it is typical of other realizations.
The numbers on the graph refer to the
values of $r_s$. The solid lines represent
singlets and the crosses triplets. The lines towards the bottom right
refer to the upper band, which results from on-site repulsion.
Features worth noticing are \textit{(i)} As expected,
spin plays no
role deep in the localized regime (since the exchange energy is proportional 
to $\langle \varphi_a | \varphi_b \rangle^2$). Perhaps more unexpectedly, 
spin also seems to have little importance near the crossover to
delocalization.
\textit{(ii)} Configurations with doubly-occupied sites play little
or no role near
the crossover -- they
become important only for $r_s \sim 2$.
\textit{(iii)} In
the localized extreme (\textit{i.e.} $r_s=12$), the rather short span
in energy of the eigenstates comes from the random and Coulomb energies;
as $r_s$ decreases,
the broadening and shift of the range of eigenergies is
attributable to the growing off-diagonal energies of $H$.

Examining $\lambda (E)$ we observe $\lambda$ to decrease
sharply with increasing $E$ for large $r_s$. This is easily understandable as 
an increase in the repulsion energy with decreasing $\lambda$. The dependence of
$\lambda$ on $E$ weakens as $r_s$ decreases and 
configuration-mixing takes place -- $\lambda$
becomes nearly independent of $E$ for $r_s=5$. The crossover takes place around
$r_s=5 \textrm{-} 7$, which is somewhat lower than the crossover in fig. 1.
The value of $\lambda$ (for $r_s=5$) is reasonably close to $L/2$,
the maximum possible separation (except in corners) when cyclic
boundary conditions are used. This provides evidence of
collective delocalization \cite{Pollak}. While the large off-diagonal energy
shows the importance of elastic hopping, the persistently large \textit{e-e}
separation shows that the electrons move in a correlated fashion
to stay apart. Studies of
$\lambda (E)$ in non-interacting systems show a lack of functional dependence,
with $\lambda$ varying at random over a large range of values.

\vspace{7 mm}
\includegraphics[9mm,45mm][70mm,115mm]{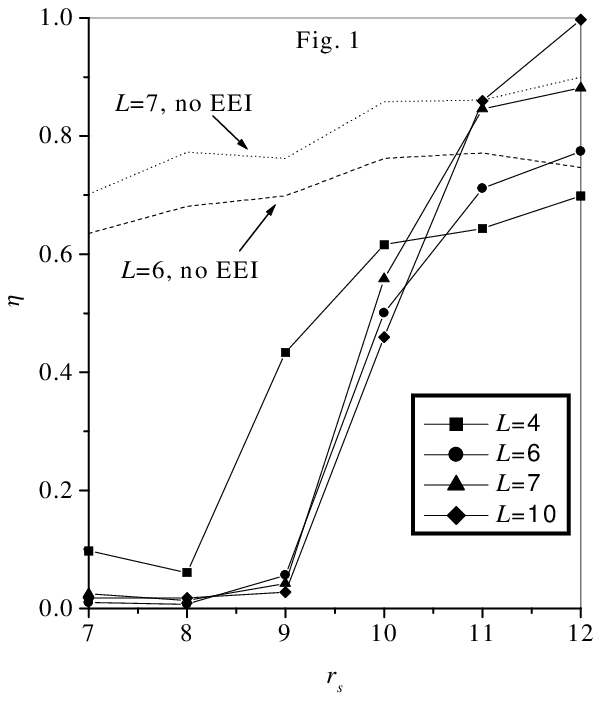}
\includegraphics[9mm,45mm][70mm,115mm]{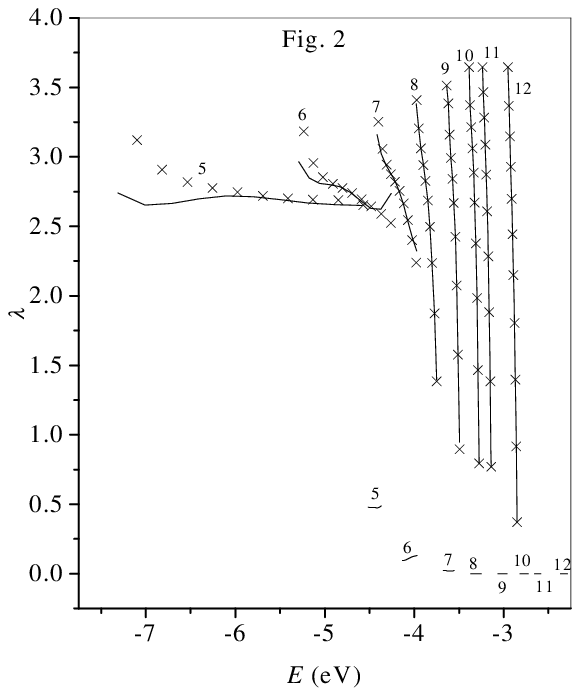}

\small
Figure 1. $\eta$ \textit{vs.} $r_s$ for $L=4,6,7,10$.
For comparison, two plots ($L=6,7$)
have been included for which the EEI is absent.

Figure 2. $\lambda$ \textit{vs.} $E_I$ for $L=7$ and a given realization
of disorder. Solid lines are fits for mixtures of singlets (lower bands
correspond to doubly-occupied sites), and crosses are fits for mixtures of
triplets. The numbers indicate the corresponding values of $r_s$.
\normalsize
\vspace{2 mm}

\textbf{Conclusions.}
The model used here yields zero density in the thermodynamic limit, and so
no definite
claims or comparisons with experiments can be made; nevertheless, it is interesting
that even this model gives 
electronic delocalization for densities $\sim 10^{12} \textrm{ cm}^{-2}$, which
is only an order of magnitude off the critical density observed ($10^{11}
\textrm{ cm}^{-2}$) in Si-MOSFETs \cite{K&DasS}. Possibly a different
choice of $W$ might give a better value for the critical density; nevertheless, 
many-electron effects probably also play an important role in delocalization
at finite electron densities.

Where collective hopping of the two electrons is coherent ($r_s<6$),
$\lambda$ can be interpreted as a coherence length.
It is of interest to note that a
crossover of $p(s)$ from Poisson to Wigner occurs at a somewhat larger $r_s$
than the crossover from a large variation in $\lambda (E)$ to $\lambda (E) \sim
\mathrm{const.}$
This suggests that (unlike in the non-interacting case) for interacting systems, delocalization
in \textit{real space} requires a somewhat larger overlap for \textit{n-n} sites than does the
transition to Wigner statistics. We do not yet understand this result well, but note
that it is in keeping with previous work \cite{Vancouver-Jerusalem}. There it was observed
that the wavefunctions are ``swiss cheese-like" without EEI, but space-filling with the
interactions. Thus, while the EEI may make the wavefunction extend over more sites, it
does not similarly increase its \textit{spatial} extent; therefore $\lambda$ might
require a larger overlap than $p(s)$ (for a
crossover to take place) because the EEI makes the wavefunction more
compact in real space.

\vspace{2 mm}
\textbf{Acknowledgements.} J. T. gratefully acknowledges the support of the
National Science Foundation under grant DMR-9803686. I. V. acknowledges
the support of the Hungarian Research Foundation OTKA (T029813, T024136,
F024135), and the A. v. Humboldt Foundation.
%


\end{document}